\documentclass[article,twocolumn]{revtex4}
\usepackage{amsfonts}
\usepackage{epsfig}
\usepackage{color}
\usepackage{graphicx}
\usepackage{dcolumn}
\usepackage{bm}

\begin{document}
\title{Rotationally induced vortices in optical cavity modes}
\author{Steven~J.~M.~Habraken}
\email{habraken@molphys.leidenuniv.nl}
\affiliation{Leiden Institute of Physics, Leiden University, P.O.
Box 9504, 2300 RA Leiden, The Netherlands}
\author{Gerard Nienhuis}
\affiliation{Leiden Institute of Physics, Leiden University, P.O.
Box 9504, 2300 RA Leiden, The Netherlands}

    \begin{abstract}
We show that vortices appear in the modes of an astigmatic optical cavity when it is put into rotation about its optical axis. We study the properties of these vortices and discuss numerical results for a specific realization of such a set-up. Our method is exact up to first order in the time-dependent paraxial approximation and involves bosonic ladder operators in the spirit of the quantum-mechanical harmonic oscillator.
    \end{abstract}

\maketitle

\section{Introduction}
Rotation is a very natural source of vorticity: when a glass of water is stirred, a vortex appears at the center. Such a vortex is a singularity of the water current and it is of the same type as the vortex that appears above a sink. When a glass of water is put into uniform rotation, for instance by placing it on a turntable, the water current vanishes at the rotation axis but is not singular. It has been known for over half a century that this is different in case of a superfluid \cite{Osborne50}. Due to the zero viscosity of, and the vortex quantization in, a superfluid, uniform rotations induce a regular pattern of equally charged vortices. Vortices appear only if the rotation frequency exceeds a certain critical value and their number increases if the rotation frequency is further increased. Eventually rotation may destabilize the superfluid. More recently, similar experiments have been performed with Bose-Einstein condensates of dilute gasses both by optically stirring the condensate \cite{Madison00} and by trapping it in a rotating elliptical potential \cite{Hodby01}.

During the past decades, optical vortices and their propagation have attracted a significant amount of attention \cite{Nye74, Indebetouw93, Rozas97, Freund00, O'Holleran08}. An optical vortex is a singularity of the phase of an optical beam and is characterized by its position in the transverse plane, its topological charge and its morphology. The vortex charge is determined by the total phase change $2\pi q$ along a contour around the vortex center and must be integer for reasons of continuity. As opposed to the vortices that appear in superfluids and Bose-Einstein condensates, optical vortices can be elliptical; their morphology is characterized by the partial derivatives of the beam profile close to the vortex center and can be represented by a point on a sphere \cite{Roux03}. The polar angle on the sphere determines the degree of ellipticity while the azimuth angle fixes the orientation in the transverse plane. The interplay between astigmatism and the propagation of optical vortices may give rise to very rich behaviour \cite{Bekshaev04, Visserthesis, Singh07}. The dynamics of optical vortices in a laser cavity has also been studied \cite{Staliunas99}.

In view of the recent interest in effects of rotating elements on optical beams \cite{Courtial98} and the physical properties of rotating mode patterns \cite{Bekshaev05}, the above-mentioned examples of rotationally-induced vortices in material systems raise the question if and how rotation induces vorticity in light fields. In this paper we address this topic by studying the optical properties of a two-mirror cavity that is put into rotation about its optical axis. This set-up is schematically drawn in Fig \ref{cavity}. We expect an effect of rotation only if the cavity lacks rotational symmetry. In analogy with the rotating elliptical potential in which Bose-Einstein condensates can be trapped, we break the rotational symmetry by taking at least one of the mirrors cylindrical or astigmatic. In the absence of rotation, such a cavity has astigmatic Hermite-Gaussian modes \cite{Siegman}, which have lines of zero intensity (line dislocations) in the transverse plane. We show that rotation deforms the cavity modes into generalized Gaussian modes \cite{Abramochkin04} and that the line dislocations are deformed into optical vortices (point singularities in the transverse plane). We study the properties of these rotationally-induced optical vortices.

This paper is organized as follows. In the next section we briefly review the propagation of optical fields through time-dependent systems and focus on the specific case of a rotating astigmatic cavity. We discuss the ladder-operator method that we apply to obtain explicit expressions of the rotating cavity modes \cite{Habraken08}. In the third section we use the approach that is discussed in Ref. \cite{Visser04} to characterize the degrees of freedom associated with the astigmatism and vorticity of the rotating cavity modes. We apply the analogy with the quantum-mechanical harmonic oscillator to derive analytical expressions of the rotating cavity modes. These are used to discuss some general properties of the vortices that appear in the modes. In the fourth section we show and discuss some numerical results for a specific realization of a uniformly rotating two-mirror cavity.

\section{Paraxial wave optics between rotating mirrors}
\subsection{Mode propagation in rotating cavity}
The mathematical description of the propagation of light through optical systems simplifies significantly if the paraxial approximation is applied. This approximation is almost always justified in experimental set-ups with optical beams. In the present case of a rotating cavity, we must account for effects that arise from the time dependence of the mirror settings. Assuming that the rotation frequency $\Omega$ is much smaller than the optical frequency $\omega$, we use the generalization of the paraxial approximation to the time-dependent case \cite{Deutsch91}. In this approximation, the electric field is purely transverse. For a propagating mode it can be written as
\begin{equation}\label{Efield}
\mathbf{E}(\mathbf{r},t)=\mathrm{Re}\left\{E_{0}\epsilon
u(\mathbf{r},t)e^{ikz-i\omega t}\right\}\;,
\end{equation}
where $E_{0}$ is the amplitude of the field, $\epsilon$ is the transverse polarization, $k$ is the wave number and $\omega=ck$ is the
optical frequency with $c$ the speed of light. The complex scalar profile $u(\mathbf{r},t)$ characterizes the large-scale spatial structure and slow temporal variations of the field. It obeys the time-dependent paraxial wave equation
\begin{equation}\label{tdpweu}
\left(\nabla_{\rho}^{2}+2ik\frac{\partial}{\partial
z}+\frac{2ik}{c}\frac{\partial}{\partial
t}\right)u(\rho,z,t)=0\;,
\end{equation}
with $\rho=(x,y)$ and $\nabla^{2}_{\rho}=\partial^{2}/\partial x^{2}+\partial^{2}/\partial y^{2}$ is the transverse Laplacian. If we omit the derivative with respect to time, this equation reduces to the standard paraxial wave equation. This equation has the same form as the Schr\"odinger equation for a free particle in 2D+1, $z$ playing the role of time. It describes the diffraction of a freely propagating stationary paraxial beam. The time derivative in Eq. (\ref{tdpweu}) accounts for the time dependence of the profile and incorporates retardation between distant transverse planes.

In addition to diffraction, the propagation of the light inside a cavity is governed by the boundary condition that the electric field must vanish on the mirror surfaces. In case of a rotating cavity this boundary condition is explicitly time dependent. A natural way to eliminate this time dependence is by transforming to a co-rotating frame, where it is sufficient to consider the behaviour of time-independent propagating modes $v(\mathbf{r})$. The transformation that connects the profile in the rotating frame to the profile in the stationary frame can be expressed as
\begin{equation}\label{rotmode}
u(\rho,z,t)=\hat{U}_{\mathrm{rot}}(\Omega t)v(\rho,z)\;,
\end{equation}
where $\Omega$ is the rotation frequency and $\hat{U}_{\mathrm{rot}}(\alpha)=\exp(-i\alpha\hat{L}_{z})$ is the operator that rotates a scalar function over an angle $\alpha$ about the $z$ axis with $\hat{L}_{z}=-i(x\partial y-y\partial x)$ the $z$-component of the orbital angular momentum operator. Substitution of the rotating mode (\ref{rotmode}) in the time-dependent paraxial wave equation (\ref{tdpweu}) yields the wave equation for $v(\rho,z)$
\begin{equation}\label{tdpwev}
\left(\nabla_{\rho}^{2}+2ik\frac{\partial}{\partial
z}+\frac{2\Omega k}{c}\hat{L}_{z}\right)v(\rho,z)=0\;
\end{equation}
In the rotating frame, the retardation term is replaced by a Coriolis term, which is familiar from particle mechanics. Since $\nabla_{\rho}$ and $\hat{L}_{z}$ commute, the formal solution of the paraxial wave equation in the rotating frame (\ref{tdpwev}) can be expressed as
\begin{equation}\label{rotprop}
v(\rho,z)=\hat{U}_{\mathrm{f}}(z)\hat{U}_{\mathrm{rot}}\left(-\frac{\Omega
z}{c}\right)v(\rho,0)\equiv\hat{U}(z)v(\rho,0)\;,
\end{equation}
where $\hat{U}_{\mathrm{f}}(z)=\exp\big(\frac{iz}{2k}\nabla_{\rho}^{2}\big)$ is the propagator corresponding to the time-independent paraxial wave equation and describes free propagation of a paraxial beam in a stationary frame. The operator $\hat{U}(z)$ has the significance of the propagator in the rotating frame. The rotation operator arises from the Coriolis term in Eq. (\ref{tdpwev}) and gives the propagating modes a twisted nature.

\begin{figure}
\begin{center}
\includegraphics[scale=0.6]{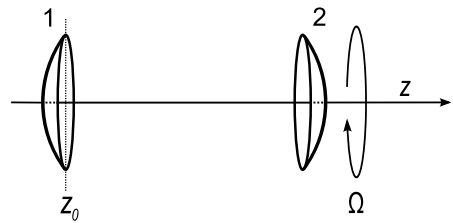}
\end{center}
\caption{\label{cavity} Schematic plot of the set-up that we study in this paper: an astigmatic two-mirror cavity that is put into rotation about its optical axis. The rotation frequency is denoted $\Omega$ and $z_{0}$ indicates the transverse reference plane.}
\end{figure}

\subsection{The modes of a rotating cavity}
In a stationary cavity, modes are usually defined as stationary solutions of the time-dependent paraxial wave equation (\ref{tdpweu}) that vanish on the mirror surfaces \cite{Siegman}. The transverse profiles of these modes are reproduced after each round trip up to a Gouy phase factor $\exp(-i\chi)$, which determines the resonant wave numbers. From Eq. (\ref{Efield}) it follows that the electric field picks up a phase $2kL-\chi$, where $L$ is the mirror separation, after each round trip so that the resonance condition reads $2kL-\chi=2\pi q$ with $q\in\mathbb{Z}$. The necessary and sufficient condition for a stationary cavity to have long-lived modes is that it is geometrically stable, i.e. that the round-trip Gouy phases $\chi$ are real so that the magnification $|\exp(-i\chi)|$ is equal to $1$.

Since a rotating cavity is time-dependent, we cannot expect time-independent modes in this case. A natural and continuous generalization of the mode criterion to the rotating case is to require that the modes adopt the time-dependence of the cavity, i.e. that they rotate along with the mirrors. These modes are time-independent in the co-rotating frame so that they obey Eq. (\ref{tdpwev}) and vanish on the mirror surfaces. Though the interplay between rotation and stability gives rise to surprisingly rich behaviour, geometrically stable rotating two-mirror cavities exist \cite{Habraken082A}. In a recent paper we have developed an analytical-algebraic method to find explicit expressions of these modes \cite{Habraken08}. The method involves two pairs of bosonic ladder operators in the spirit of the quantum-mechanical harmonic oscillator that generate a complete and orthogonal set of cavity modes. In a given transverse reference plane, which we take close to the first mirror and denote by $z=z_{0}$ as is indicated in Fig. \ref{cavity}, the profiles of the cavity modes can be expressed as
\begin{equation}\label{modes}
v_{nm}(\rho,z_{0})=\frac{1}{\sqrt{n!m!}}\left(\hat{a}^{\dag}_{1}(z_{0})\right)^{n}\left(\hat{a}^{\dag}_{2}(z_{0})\right)^{m}v_{00}(\rho,z_{0})\;,
\end{equation}
where $\hat{a}^{\dag}_{1}(z_{0})$ and $\hat{a}^{\dag}_{2}(z_{0})$ are the two raising operators in the reference plane. The fundamental mode $v_{00}(z_{0})$ is fixed up to a phase factor by the requirement that acting on it with the corresponding lowering operators must give zero, i.e. $\hat{a}_{1}(z_{0})v_{00}(z_{0})=\hat{a}_{2}(z_{0})v_{00}(z_{0})=0$. The ladder operators are linear combinations of the position operators $\hat{\rho}=(x,y)$ and the conjugate momentum operators $k\hat{\theta}=-i\nabla_{\rho}$ with $k$ the wave number of the mode. The expectation values $<v|\hat{\rho}|v\rangle$ and $<v|\hat{\theta}|v\rangle$ have the significance of the average transverse position and the average propagation direction of the beam. The position and propagation direction operators obey canonical commutation relations $[\hat{\rho}_{i},k\hat{\theta}_{j}]=i\delta_{ij}$, where the indices $i$ and $j$ run over the $x$ and $y$ and with $\delta_{ij}$ the Kronecker delta function. In the reference plane $z_{0}$ the lowering operators can be expressed as
\begin{equation}\label{lowop}
\hat{a}_{i}(z_{0})=\sqrt{\frac{k}{2}}\left(t_{i}(z_{0})\hat{\rho}-r_{i}(z_{0})\hat{\theta}\right)\;,
\end{equation}
where the index $i$ runs over 1 and 2. The complex vectors $t_{i}(z_{0})$ and $r_{i}(z_{0})$ have two components and are chosen such that $\mu_{i}=(r_{i},t_{i})$ is an eigenvector of the round-trip ray matrix in the co-rotating frame. In the present case of an astigmatic optical cavity, this is a real $4\times 4$ matrix, which can be expressed as
\begin{equation}\label{Mrt}
M_{\mathrm{rt}}=M_{1}\cdot M(L)\cdot M_{2}\cdot M(L)\;,
\end{equation}
where $M(z)=M_{\mathrm{f}}(z)\cdot M_{\mathrm{rot}}(-\Omega z/c)$ is the ray matrix that corresponds to $\hat{U}(z)$ and describes free propagation in the co-rotating frame, $M_{\mathrm{f}}(z)$ is the ray matrix that describes propagation in a stationary frame and $M_{\mathrm{rot}}(\alpha)$ is the ray matrix that describes a rotation over an angle $\alpha$ in the transverse plane. The ray matrices $M_{1,2}$ describe the mirrors $1$ and $2$ respectively and are fully determined by their radii of curvature and orientation in the transverse plane. These ray matrices are generalizations of the standard $2\times 2$ ray matrices, which can be found in any textbook of optics, to the astigmatic case. Explicit expressions of the $4\times 4$ ray matrices are given in Ref. \cite{Habraken08}.

As opposed to the unitary propagator $\hat{U}(z)$ and the rotation operator $\hat{U}_{\mathrm{rot}}(\alpha)$, which act in the Hilbert space of transverse modes, ray matrices describe real linear transformations in the transverse phase space $(\rho,\theta)$. Formally speaking, this phase space is a symplectic manifold and the real and linear transformations that preserve its mathematical structure form the real symplectic group $Sp(\mathbb{R},4)$. Symplectic groups and various physically relevant aspects of symplectic geometry have been studied in detail, see for instance Ref. \cite{Guilleman84}. The ladder operators act in the mode space, but since they are constructed from the ray vectors $\mu_{i}=(r_{i},t_{i})$ and transform accordingly, the algebraic properties of the round-trip ray matrix (\ref{Mrt}) are essential for the ladder operator approach to be applicable. From these properties, it follows that its eigenvalues are either real or pairwise complex conjugate phase factors \cite{Habraken07}. The rotating cavity is geometrically stable only in the latter case. In this case, the corresponding pairwise complex conjugate eigenvectors $\mu_{1,2}=(r_{1,2},t_{1,2})$ and $\mu^{\ast}_{1,2}=(r^{\ast}_{1,2},t^{\ast}_{1,2})$ can be chosen such that
\begin{equation}\label{rtprop1}
r_{i}(z_{0})t_{j}(z_{0})-r_{i}(z_{0})t_{j}(z_{0})=0\;,
\end{equation}
and
\begin{equation}\label{rtprop2}
r^{\ast}_{i}(z_{0})t_{j}(z_{0})-r_{i}(z_{0})t^{\ast}_{j}(z_{0})=2i\delta_{ij}\;,
\end{equation}
where the indices $i$ and $j$ take the values 1 and 2. The complex conjugate eigenvectors $\mu^{\ast}_{1,2}$ generate the raising operators $\hat{a}^{\dag}_{1,2}$ according to Eq. (\ref{lowop}) and the special properties of the eigenvectors (\ref{rtprop1}, \ref{rtprop2}) guarantee that the ladder operators obey bosonic commutation relations
\begin{equation}\label{boscom}
[a_{i}(z_{0}),a_{j}(z_{0})]=0\quad\mathrm{and}\quad[\hat{a}_{i}(z_{0}),\hat{a}^{\dag}_{j}(z_{0})]=\delta_{ij}\;,
\end{equation}
where the indices $i$ and $j$ run over 1 and 2, so that the modes (\ref{modes}) form a complete and orthonormal set in the reference plane.

The four eigenvalues of the round-trip ray matrix (\ref{Mrt}) can be written as $\exp(i\chi_{1,2})$ and $\exp(-i\chi_{1,2})$, where $\chi_{1,2}$ are the round-trip Gouy phase that are picked up by the lowering operators. They determine the resonant wave numbers for the $(n,m)$ cavity mode according to
\begin{equation}\label{rescond}
2kL-\chi_{1}\left(n+\frac{1}{2}\right)-\chi_{2}\left(m+\frac{1}{2}\right)=2\pi q\;,\quad q\in\mathbb{Z}\;.
\end{equation}

The ladder operators that generate the rotating cavity modes in an arbitrary transverse plane can be constructed according to Eq. (\ref{lowop}) and its hermitian conjugate by using that the ray vectors $\mu_{i}=(r_{i},t_{i})$ transform according to
\begin{equation}
\left(\!\begin{array}{c}r_{i}(z)\\t_{i}(z)\end{array}\!\right)=M(z-z_{0})\left(\!\begin{array}{c}r_{i}(z_{0})\\t_{i}(z_{0})\end{array}\!\right)
\end{equation}
The properties given by Eqs. (\ref{rtprop1}) and (\ref{rtprop2}) are preserved under this transformation so that the ladder operators obey the bosonic commutation relations (\ref{boscom}) in all transverse planes. So far, we have considered only the mode that propagates from left to right (from mirror 1 to mirror 2 in Fig. \ref{cavity}). By using the ray matrix that describes one of the mirrors (say mirror 2), one can construct the ladder operators that generate the modes that propagate in the opposite direction: $(r^{\leftarrow}_{i}(z),t^{\leftarrow}_{i}(z))= M(z_{0}+L-z)M_{2}M(z_{0}+L-z)(r^{\rightarrow}_{i}(z),t^{\rightarrow}_{i}(z))$. The actual field inside the cavity is a linear combination of the left and right propagating modes and can be expressed as
\begin{eqnarray}\label{cavityfield}
\mathbf{E}_{nm}(\mathrm{r},t)=\mathrm{Re}\Big\{-iE_{0}\epsilon\big(v_{nm}^{\rightarrow}(\rho,z)e^{ikz-i\omega t}-\qquad\nonumber\\ \quad v_{nm}^{\leftarrow}(\rho,z)e^{-ikz-i\omega t}\big)\Big\}\;,
\end{eqnarray}
where the minus sign accounts for the fact that a mode changes sign when it is reflected by a mirror. In the rest of this paper, we focus on the mode profile $v_{nm}$ rather than on the actual cavity field (\ref{cavityfield}), since this is the profile that would be measured in any realistic experiment. The expressions of modes that we have given in this section hold in the co-rotating frame, the corresponding expressions of the rotating mode patterns $u_{nm}(\rho,z,t)$ in the stationary frame can be obtained by applying Eq. (\ref{rotmode}).

\begin{figure}
\begin{center}
\includegraphics[scale=0.6]{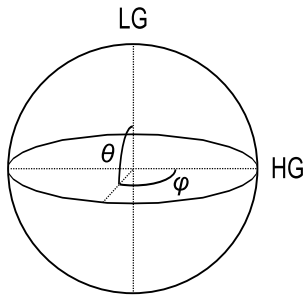}
\end{center}
\caption{\label{HLsphere}
The Hermite-Laguerre sphere, which describes the degrees of freedom associated with the nature of higher-order paraxial optical modes. Each point $(\phi_{\mathrm{HL}},\theta_{\mathrm{HL}})$ corresponds to a complete and orthonormal set of modes; the poles ($\theta_{\mathrm{HL}}=0,\pi$) correspond to Laguerre-Gaussian modes while points on the equator ($\theta_{\mathrm{HL}}=\pi/2$) correspond to Hermite-Gaussian modes. Intermediate values of the polar angle $\theta_{\mathrm{HL}}$ give rise to generalized Gaussian modes. The azimuth angle $\phi_{\mathrm{HL}}$ determines the orientation of the higher-order mode patterns in the transverse plane.}
\end{figure}

\begin{figure}[!t]
\begin{center}
\includegraphics[scale=0.6]{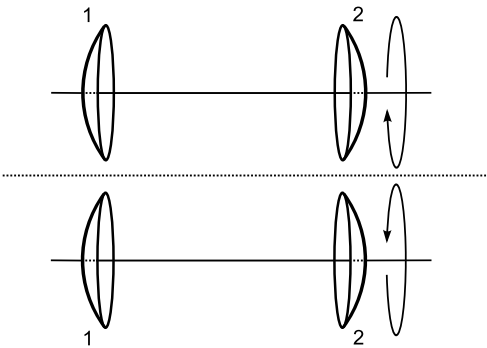}
\end{center}
\caption{\label{symmetry} A stationary cavity has inversion symmetry so that it is equivalent to its mirror image (the dashed line indicates the mirror plane). Since the rotation direction changes under inversion, this is not true of a rotating cavity.}
\end{figure}

\section{Ladder operators and vortices}
\subsection{Analytical expressions of the modes}
In order to give more explicit expressions of the rotating cavity modes (\ref{modes}), we combine the vectors $r_{1,2}$ and $t_{1,2}$ in two matrices, which are defined as
\begin{equation}
\mathsf{R}(z)=-i\left(\begin{array}{c}r_{1}(z)\\r_{2}(z)\end{array}\right)\qquad\mathrm{and}\qquad
\mathsf{T}(z)=\left(\begin{array}{c}t_{1}(z)\\t_{2}(z)\end{array}\right)\;,
\end{equation}
where the factor $-i$ is introduced for notational convenience. The special properties (\ref{rtprop1}, \ref{rtprop2}) can be summarized as
\begin{equation}\label{RTprop}
\mathsf{R}^{\mathrm{T}}\mathsf{T}-\mathsf{T}^{\mathrm{T}}\mathsf{R}=0\quad\mathrm{and}\quad \mathsf{R}^{\dag}\mathsf{T}+\mathsf{T}^{\dag}\mathsf{R}=2
\end{equation}
As was mentioned already, the fundamental mode $v_{00}(\rho,z)$ is fixed by the requirement that acting on it with both lowering operators must give 0. The two first order differential equations that are thus obtained have the solution
\begin{equation}\label{v00}
v_{00}(\rho,z)=\sqrt{\frac{k}{\pi\det\mathsf{R}(z)}}\exp\left(-k\rho^{\mathrm{T}}\mathsf{S}(z)\rho/2\right)\;,
\end{equation}
where $\mathsf{S}=\mathsf{R}^{-1}\mathsf{T}$ is a $2\times 2$ matrix. It makes sense to decompose it into its real and imaginary parts $\mathsf{S}=\mathrm{S}_{\mathrm{r}}+i\mathrm{S}_{\mathrm{i}}$. The matrices $\mathrm{S}_{\mathrm{r}}$ and $\mathrm{S}_{\mathrm{i}}$ respectively characterize the astigmatism of the Gaussian intensity and phase patterns. From the properties of $\mathsf{R}$ and $\mathsf{T}$ (\ref{RTprop}) it can be shown easily that $\mathsf{S}$ is symmetric and that its real part $\mathrm{S}_{\mathrm{r}}$ is positive definite so that the mode profile is square-integrable. The profile in Eq. (\ref{v00}) has been normalized properly. In order to characterize the other degrees of freedom, it is convenient to rewrite the $z$-dependent lowering operators (\ref{lowop}) as
\begin{equation}
\left(\!\begin{array}{c}\hat{a}_{1}(z)\\ \hat{a}_{2}(z)\end{array}\!\right)=\sqrt{\frac{k}{2}}\mathsf{R}(z)\left(\mathsf{S}(z)\hat{\rho}+i\hat{\theta}\right)\;.
\end{equation}
By using Eq. (\ref{RTprop}) we find that $\mathsf{R}^{\dag}\mathsf{R}=\mathsf{S}_{\mathrm{r}}^{-1}$ so that $\mathsf{R}(z)$ can be written as $\sigma(z)\mathsf{S}(z)_{\mathrm{r}}^{-1/2}$, where $\sigma$ is a unitary $2\times 2$ matrix. Notice that $\mathsf{S}_{\mathrm{r}}$ is real, symmetric and positive definite so that $\mathsf{S}_{\mathrm{r}}^{-1/2}$ is well-defined. By making use of the operator identity $e^{A}Be^{-A}=B+[A,B]+\frac{1}{2!}[[A,[A,B]]+...$ and the canonical commutation relations $[\hat{\rho},k\hat{\theta}]=i\delta_{ij}$ we find that
\begin{equation}
i\mathsf{S_{\mathrm{i}}}\hat{\rho}+i\hat{\theta}=e^{-ik\rho\mathsf{S}_{\mathrm{i}}\rho/2}\hat{\theta}e^{ik\rho\mathsf{S}_{\mathrm{i}}\rho/2}
\end{equation}
so that the $z-$dependent lowering operators can be expressed as
\begin{equation}\label{lowop2}
\left(\!\begin{array}{c}\hat{a}_{1}\\ \hat{a}_{2}\end{array}\!\right)=\sqrt{\frac{k}{2}}\sigma e^{-ik\rho\mathsf{S}_{\mathrm{i}}\rho/2}\left(\mathsf{S}^{1/2}_{\mathrm{r}}\hat{\rho}+ i\mathsf{S}^{-1/2}_{\mathrm{r}}\hat{\theta}\right)e^{ik\rho\mathsf{S}_{\mathrm{i}}\rho/2}\;.
\end{equation}
By introducing real scaled coordinates $\rho'=\sqrt{k}\mathsf{S}_{\mathrm{r}}^{1/2}\rho=(x',y')$ that account for the astigmatism of the intensity pattern, the product of $\sqrt{k/2}$ and the linear combination of $\hat{\rho}$ and $\hat{\theta}$ between the brackets takes the form of the lowering operators of a dimensionless quantum-mechanical harmonic oscillator in 2D+1. From right to left, both the $z$-dependent lowering operators (\ref{lowop2}) and the corresponding raising operators $\hat{a}^{\dag}_{1,2}$ first remove the curved wave front, then modify the mode pattern and eventually restore the wave front again. The unitary matrix $\sigma$ describes the additional degrees of freedom, which characterize the nature of the higher-order modes. Overall phase factors in the rows of $\sigma$ are physically irrelevant and without loss of generality we can fix its determinant such that $\sigma\in SU(2)$. Since $\sigma$ is unitary, its rows are not independent and we can write
\begin{equation}\label{sigma}
\sigma=\left(\begin{array}{cc}\sigma_{x}&\sigma_{y}\\-\sigma^{\ast}_{y}&\sigma^{\ast}_{x}\end{array}\right)\;,
\end{equation}
with $|\sigma_{x}|^{2}+|\sigma_{y}|^{2}$. In complete analogy with the Poincar\'e sphere for polarization vectors (or the Bloch sphere for spin-1/2 states) the degrees of freedom associated with the complex vector $(\sigma_{x},\sigma_{y})$ can be mapped onto the so-called Hermite-Laguerre sphere \cite{Visser04}. As is indicated in Fig. \ref{HLsphere}, every point on the sphere corresponds to two pairs of bosonic ladder operators that generate a complete set of higher-order modes. In this respect, it is different from the Poincar\'e and Bloch spheres, on which every point corresponds to a single state. The poles on the Hermite-Laguerre sphere correspond with $(\sigma_{x},\sigma_{y})=(1,\pm i)/\sqrt{2}$ and yield ladder operators that generate Laguerre-Gaussian modes. Points on the equator correspond to $(\sigma_{x},\sigma_{y})=(\cos\phi,-\sin\phi)$ and give rise to Hermite-Gaussian modes. Intermediate values of the polar angle $0<\theta_{\mathrm{HL}}<\pi/2$ correspond to generalized Gaussian modes \cite{Abramochkin04}. The azimuth angle $\phi_{\mathrm{HL}}$ fixes the orientation of the higher-order mode patterns in the transverse plane. The rows of $\sigma$ (\ref{sigma}) correspond to antipodal points on the sphere so that it is, strictly speaking, sufficient to consider one of the hemispheres only. In general, the separation of the degrees of freedom in terms of symmetric matrix $\mathsf{S}$, which characterizes the astigmatism, and the unitary matrix $\sigma$, which determines the nature of the higher order modes, is local in the sense that only holds in a single transverse plane $z$. The evolution of $\sigma$ under propagation and optical elements depends on the astigmatism and vice versa.

Using the defining identity of the Hermite polynomials $H_{n}(x)\exp(-x^{2}/2)=(x-\frac{\partial}{\partial x})^{n}\exp(-x^{2}/2)$ and the binomial expansion $(a+b)^{n}=\sum_{p=0}^{n}{n \choose p}a^{p}b^{n-p}$ for $[a,b]=0$ the higher order modes can be expressed as
\begin{eqnarray}\label{vnm}
v_{nm}(\rho',z)=e^{-\rho'^{2}/2-i\rho'\mathsf{S}_{\mathrm{r}}^{-1/2}\mathsf{S}_{\mathrm{i}}\mathsf{S}_{\mathrm{r}}^{-1/2}\rho'/2}\times \qquad\qquad\qquad\nonumber\\
\sum_{p=0}^{n}\sum_{q=0}^{m}{n \choose p}{m \choose q}\left(\sigma^{\ast}_{x'}\right)^{p}\left(\sigma_{y'}\right)^{q} \left(\sigma^{\ast}_{y'}\right)^{n-p}\left(\sigma_{x'}\right)^{m-q}\times\quad\nonumber\\H_{k+l}(x')H_{n+m-k-l}(y')\;,\quad
\end{eqnarray}
where $\rho'=(x',y')$ are the scaled coordinates that account for the astigmatism of the intensity pattern. The above expression holds in the co-rotating frame. The corresponding expressions for the rotating modes $u_{nm}(\rho',z,t)$ in the stationary frame, can be obtained by applying Eq. (\ref{rotmode}).

\subsection{Vortices in higher order modes}
As is indicated in Fig. \ref{symmetry}, a stationary two-mirror cavity has inversion symmetry so that it must have the same modes as its mirror image. In the non-degenerate case, this implies that the mode profiles close to the mirrors are real apart from the overall curved wave fronts \cite{Habraken07}. Since this symmetry property holds for modes of all order, it implies that the higher-order modes of a stationary cavity close to the mirrors must be Hermite-Gaussian. Hermite-Gaussian modes do have line dislocations (lines along which the phase jumps by $\pi$) in the transverse plane, but do not have vortices (phase singularities).

Rotation obviously breaks the inversion symmetry of a stationary cavity. As a result, the modes of a rotating cavity have additional phase structure. The fundamental mode attains a twist; though the electric field vanishes on the mirror surfaces, its wave fronts do not fit the local curvature of the mirrors. From Eq. (\ref{v00}), it is clear that no vortices can appear in the fundamental mode. Vortices appear as zeros of the polynomial part of the profile of the higher order modes (\ref{vnm}). Due to the rotation of the cavity, the line dislocations in the higher-order Hermite-Gaussian modes are deformed into elliptical vortices.

From the fact that $\sigma$ (\ref{sigma}) is a unitary matrix, it follows that the two raising operators $\hat{a}^{\dag}_{1,2}$ generate vortices with opposite charge. As a result, the vortices that appear at the center of the $v_{01}$ and $v_{10}$ modes have equal but opposite topological charges $\pm 1$. Since both vortices are spherical (canonical) in the scaled coordinates, their morphologies are determined only by the astigmatism of the intensity pattern. In general, the raising operators split, displace and introduce vortices in the cavity modes so that the vortex pattern in the higher-order modes (\ref{vnm}) can be very complicated. More explicit expressions have been given only in limiting cases \cite{Bekshaev04, Visserthesis}.

The fundamental mode contains only even powers of the position coordinates $\rho=(x,y)$ and is even under inversion in the transverse plane: $\rho\rightarrow-\rho$. The ladder operators are linear in the position and propagation direction operators and are obviously odd under this inversion. It follows that the rotating cavity modes (\ref{vnm}) are even or odd, depending on the parity of the total mode number $n+m$. Odd modes have a vortex at the center of the mode patterns, whereas even modes in general do not.

\begin{figure}
\begin{center}
\includegraphics[scale=1]{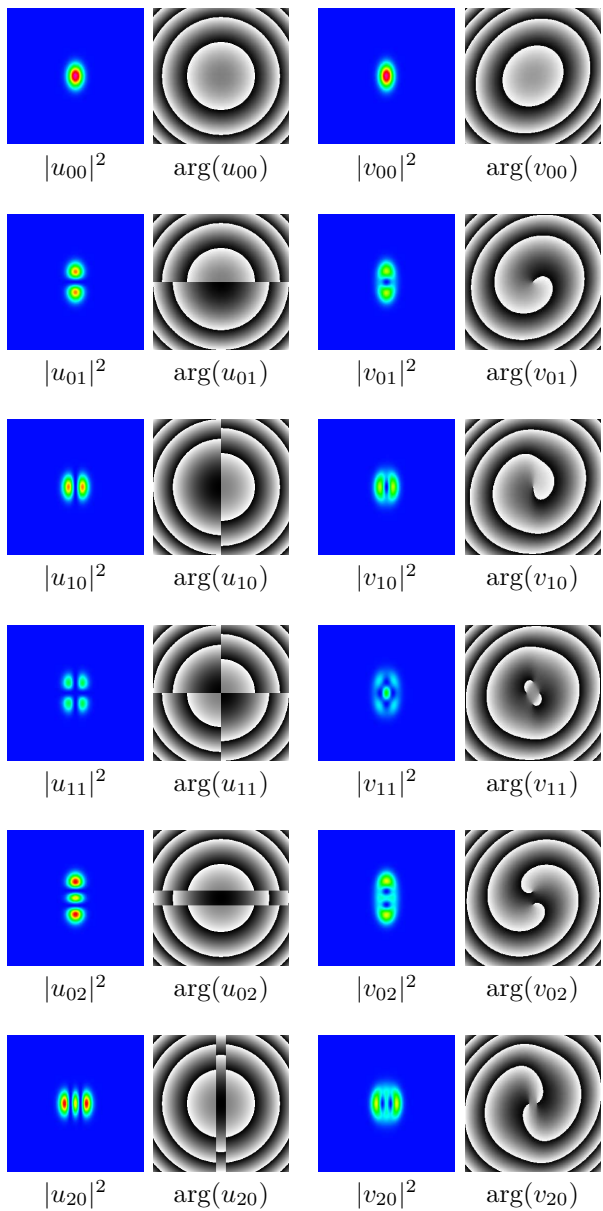}
\end{center}
\caption{\label{modeplots} False-color intensity and phase patterns of the modes of a cavity consisting of a spherical and a stationary (left) and rotating (right) cylindrical mirror. The radius of curvature of the spherical mirror is equal to $4L$, with $L$ the mirror separation. The radius of curvature of the cylindrical mirror is equal to $2L$. The plots show the mode patterns in the co-rotating frame close to the spherical mirror and the cylindrical mirror is flat in the vertical direction. In the right plots, the rotation frequency is equal to $\Omega_{0}/10$ with $\Omega_{0}=c\pi/L$.}
\end{figure}

\section{Examples}
In this section we illustrate the rotational effects on the structure of two-mirror cavity modes by further investigating a specific example. We focus on a cavity consisting of a stationary spherical and a rotating cylindrical mirror, which is the simplest realization a uniformly-rotating two-mirror cavity. The radius of curvature of the spherical mirror is taken $4L$, with $L$ the mirror separation while the radius of curvature of the cylindrical mirror is chosen $2L$. This cavity is geometrically stable for rotation frequencies up to $\Omega_{0}/6$, where $\Omega_{0}=c\pi/L$, with $c$ the speed of light, is frequency corresponding to the cavity round-trip time \cite{Habraken082A}.

In the absence of rotation, the modes of the cavity $u_{nm}$ are defined as the stationary solutions of the time-dependent paraxial wave equation (\ref{tdpweu}) that vanish on both mirror surfaces. The transverse intensity and phase patterns close to the spherical mirror of the zeroth, first and second order modes of the cavity are plotted in the left column of Fig. \ref{modeplots}. Though the astigmatism is different in other transverse planes (in particular close to the cylindrical mirror), the modes are Hermite-Gaussian in all transverse planes. Their phase structure clearly reflects the fact that the wave fronts fit the curvature of the (spherical) mirror. The phase jumps of $2\pi$, which appear along circles in the transverse plane, are an artifact of the definition of the phase of the beam and are not physical. The phase jumps of $\pi$, however, which arise from the Hermite polynomials in Eq. (\ref{vnm}) and appear along horizontal and vertical lines in the transverse plane, are physical and could be measured interferometrically.

If the cylindrical mirror is put into rotation, the mode structure changes significantly. The modes of a rotating cavity are defined as co-rotating solutions of the time-dependent paraxial wave equation (\ref{tdpweu}), or equivalently solutions of Eq. (\ref{tdpwev}), that vanish on the mirror surfaces. In the co-rotating frame, propagating modes attain a twist due to retardation, which is reflected by the Coriolis term in Eq. (\ref{tdpwev}). The effect of rotation on the intensity and phase patterns of the zeroth, first and second order cavity modes is illustrated in the right column of Fig. \ref{modeplots}, for which the rotation frequency is equal to $\Omega_{0}/10$. The mode patterns are clearly affected by rotation. The intensity patterns become more similar to Laguerre-Gaussian modes and have obviously the structure of generalized Gaussian modes. Though the electric field vanishes on the mirror surfaces, the wave front no longer fit their local curvature. This is most apparent in case of the fundamental mode $v_{00}$, where the curves of constant phase close to the spherical mirror have become elliptical. The non-parallel orientation of the elliptical intensity and phase patterns of the fundamental mode $v_{00}$ reflects its twisted nature. The higher-order modes also attain a twist. Moreover, vortices appear in their phase patterns. The results confirm that the vortices in the $v_{01}$ and $v_{10}$ have opposite charges and that only modes with odd $n+m$ have a vortex at the center. In the modes with equal mode numbers $n$ and $m$, the total vortex charge is equal to zero so that the curves of constant phase that contain all vortices are closed.

\begin{figure}[!t]
\begin{center}
\includegraphics[scale=0.8]{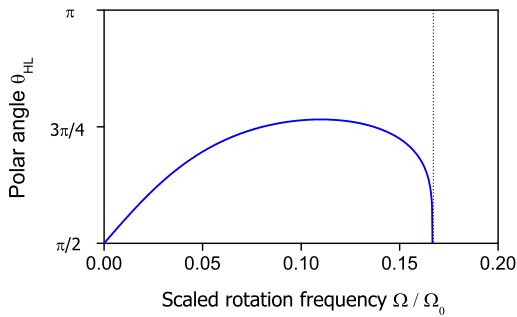}
\end{center}
\caption{\label{polar} Dependence on the rotation frequency $\Omega$ of the polar angle $\theta_{\mathrm{HL}}$ on the Hermite-Laguerre sphere for a cavity consisting of a spherical and a rotating cylindrical mirror. The characterization of the modes in terms of $(\phi_{\mathrm{HL}},\theta_{\mathrm{HL}})$ is local and the plot specifies the nature of the higher-order modes in the transverse plane close to the spherical mirror.}
\end{figure}

Due to a combined symmetry that survives rotation, the intensity patterns of the rotating modes are aligned along the axes of the cylindrical mirror while the phase patterns are not \cite{Habraken08}. As a result the azimuth angle on the Hermite-Laguerre sphere $\phi_{\mathrm{HL}}$, which specifies the orientation of the higher-order mode patterns, does not depend on the rotation frequency. This is not true for the polar angle $\theta_{\mathrm{HL}}$, which specifies whether the modes are Hermite-Gaussian, Laguerre-Gaussian or generalized Gaussian modes and is a measure of their vorticity. Its dependence on the rotation frequency $\Omega$ is shown in Fig. \ref{polar}. In general, the characterization of the cavity modes in terms of a vector on the Hermite-Laguerre sphere is local and the plot in Fig. \ref{polar} characterizes the higher-order cavity modes close the the spherical mirror. The plot confirms that, due to the fact that rotation breaks the inversion symmetry of a stationary cavity, the cavity modes are continuously deformed from Hermite-Gaussian modes in the stationary case into generalized Gaussian modes in the rotating case. For not too large values of the rotation frequency, the rotationally induced vorticity is proportional to the rotation frequency. For larger values something surprising happens: at some point, the vorticity starts to decrease with increasing rotation frequencies and eventually the modes become Hermite-Gaussian again. This is due to the fact that this cavity is destabilized by rotation at a rotation frequency $\Omega_{0}/6$. At this point, the modes lose their confinement in one of the transverse directions (in this case in the vertical direction, i.e. the direction in which the cylindrical mirror is flat) so that the elliptical vortices are stretched to become line dislocations again. Though the vorticity in the modes disappears if this transition is approached, the orbital angular momentum diverges at this point due to the diverging astigmatism \cite{Habraken082A}. In specific cases, rotation can also stabilize a cavity that is unstable in the absence of rotation. In such cases we expect the opposite behaviour. Due to the rotation of the cavity, we retrieve Hermite-Gaussian modes at the point where the cavity is stabilized, while mode deformation and induced vorticity appear for even larger values of the rotation frequency.

\section{Conclusion and outlook}

In this paper, we have applied a ladder operator method to study the vortices that appear in the modes of an astigmatic two-mirror cavity when it is put into rotation about the optical axis. The modes of a rotating cavity are defined as solutions of the time-dependent paraxial wave equation (\ref{tdpweu}) that rotate along with the cavity and vanish on the mirror surfaces. This mode criterion is a continuous generalization of the requirement that the modes of a stationary cavity are stationary solutions of Eq. (\ref{tdpweu}) that vanish on the mirror surfaces. The rotating cavity modes are stationary solutions in a co-rotating frame where mode propagation is twisted due to the finite speed of light. Due to this twist, rotation deforms the astigmatic Hermite-Gaussian modes of a stationary cavity into generalized Gaussian modes. The line dislocations in the Hermite-Gaussian modes are deformed into elliptical vortices. In a recent paper \cite{Habraken082A}, we have shown that rotation can destabilize a two-mirror cavity that is stable in the absence of rotation. When such a transition is approached, the cavity modes lose their confinement in one transverse direction so that the elliptical vortices are stretched to become line dislocations again. This is illustrated in Fig. \ref{polar}.

An interesting but open question is how rotation would affect the optical properties of a geometrically unstable cavity, especially when close to the rotationally induced transition from an unstable to a stable geometry. A geometrically unstable cavity is intrinsically lossy \cite{Siegman} and the propagation of light inside an unstable cavity is dominated by diffraction at the sharp edges of the mirrors \cite{Southwell86}. Both mathematically and physically this system is fundamentally different from the stable cavities that we have studied in this paper and it is not possible to apply or generalize our method to such a system.

As opposed to a normal fluid, a spatially confined optical cavity mode attains vorticity when it is put into uniform rotation. In this respect, it has some similarity with superfluids and Bose-Einstein condensates. Compared to those systems, however, optical beams have many more degrees of freedom to cope with rotation. The phase of Bose-Einstein condensates and superfluids is fixed by the broken gauge invariance. The only way in which such systems can attain orbital angular momentum, is by locally creating cylindrically symmetric vortices while keeping their phase fixed elsewhere. As a result, vortices appear only when the rotation frequency exceeds a certain threshold and the number of vortices increases when the rotation frequency is further increased. For optical cavity modes, on the other hand, both the astigmatism of the phase and intensity patterns and the properties of the optical vortices are affected by rotation. Both contribute to the orbital angular momentum \cite{Visser04}, which typically increases with increasing values of the rotation frequency even though the number of vortices does in general not increase.

For typical values of the mirror separation, the rotation frequency corresponding to the cavity round-trip time is in the \textit{MHz}-range. Although mechanical rotation frequencies in this range are not experimentally feasible, there are several possible routes towards experimentally realizing a set-up that captures the essential features of a rotating astigmatic two-mirror cavity. Mechanical vibration frequencies in the \textit{MHz}-range can be achieved by using a piezoelectric actuator. Developing a similar device to simulate rotations at \textit{MHz}-frequencies is challenging but, at least in principle, not impossible. Another possible way of realizing a set-up that is similar to the one that is discussed in this paper, is by applying a rotating astigmatic mode pattern, which can be constructed from its stationary Doppler-shifted frequency components, to optically induce a rotating refractive-index pattern in a material with a Kerr-nonlinearity. Such a pattern could be used in transmission to realize a two-mirror cavity with a rotating lens. The optical properties of such a cavity could be observed by using a laser at another wave length. Besides the dynamical realizations of the set-up, one could try to mimic the twisted mode propagation in the rotating frame by using image rotators. Since it is essential that back and forth propagating modes are rotated in the same direction, a two-mirror cavity with an image rotator between the mirrors does not do the job. Instead, one could use a ring resonator with four, pairwise identical, mirrors in which the light passes an image rotator after each mirror. Although the settings of the optical elements in this set-up are time-independent it does capture the optical properties of a rotating two-mirror cavity. A difficulty with this set-up is that the clockwise and counterclockwise propagating modes are frequency degenerate. In a passive set-up one of the two modes can be selected by injecting light in one direction only. In an active realization of such a set-up, the degeneracy can be lifted by applying the polarization degree of freedom.

It is a pleasure to thank Michiel~J.~A. de Dood and Eric~R. Eliel for fruitful discussions regarding possible experimental realizations of the set-up that is discussed in this paper.


\end{document}